# An Agent-based Realisation for a continuous Model Adaption Approach in intelligent Digital Twins

Daniel Dittler*, Peter Lierhammer**, Dominik Braun*,
Timo Müller*, Nasser Jazdi*, Michael Weyrich*

*Institute of Industrial Automation and Software Engineering, University of Stuttgart, Stuttgart, DE70550
Germany (Tel: +49-711-685-67321; e-mail: Daniel.Dittler@ias.uni-stuttgart.de)
**Institute of Energy Economics and Rational Energy Use, University of Stuttgart, Stuttgart, DE70550
Germany (e-mail: Peter.Lierhammer@ier.uni-stuttgart.de)

**Abstract:** The trend in industrial automation is towards networking, intelligence and autonomy. Digital Twins, which serve as virtual representations, are becoming increasingly important in this context. The Digital Twin of a modular production system contains many different models that are mostly created for specific applications and fulfil different requirements. Especially simulation models, which are created in the development phase, can be used during the operational phase for applications such as prognosis or operation-parallel simulation. Due to the high heterogeneity of the model landscape in the context of a modular production system, the plant operator is faced with the challenge of adapting the models in order to ensure an application-oriented realism in the event of changes to the asset and its environment or the addition of applications. Therefore, this paper proposes a concept for the continuous model adaption in the Digital Twin of a modular production system during the operational phase. The benefits are then demonstrated by an application scenario and an agent-based realisation.

## 1. INTRODUCTION

Globalisation, volatile markets, shorter innovation and product life cycles require increasingly flexible and adaptable systems (Järvenpää et al., 2016; Müller et al., 2020). With the rise of Industry 4.0 and the associated increase in digitalisation and networking, data, services and functions can be held and used where they provide the greatest benefit (Jazdi, 2014). This leads to a dissolution of the classic automation pyramid, which is being replaced by modular, decentrally organised systems (Schlick et al., 2013) to enable the required flexibility and adaptability. On the one hand, this offers many potentials for future automation technology, but on the other hand, it also results in new challenges. The increasing variety of distributed data, services and functionalities, where dependencies and interactions exist, leads to increasing complexity (Müller-Schloer et al., 2012). A major goal of plant operators is to run their production systems with the highest possible availability (Müller et al., 2021). Particular applications such as predictive maintenance, quality prediction or reconfiguration are becoming increasingly important in the operational phase. To realise such applications, algorithms from the fields of artificial intelligence or mathematical optimisation are used. This requires information that, in the course of the above-mentioned digitalisation, is increasingly represented in the form of models (Rosen et al., 2015). The concept of the Digital Twin based on (Ashtari et al., 2019) realises such a model-based virtual mapping of the real world into the information world and thus represents a significant advance for modelling, simulation and optimisation technologies. The Digital Twin thus offers the possibility to address the above-mentioned applications, especially on the basis of conducting simulative experiments (Dittler et al., 2022b).

During the development phase of a system, various types of models (e.g. behaviour, functional, structural models) (Oestersötebier, 2018) are developed for the intended application and operating scenarios. This leads to a very heterogeneous model landscape, because models from different disciplines (mechanics, electronics, software), from different component vendors (Scheifele, 2019) and various modelling tools come together in the context of a modular production system (Stegmaier et al., 2022a).

These different models can be related to each other by means of a Digital Twin and used, for example, for virtual commissioning, but also for the aforementioned applications in the operational phase. Changes in the requirements of an application, new applications or changes at the asset and its environment require a model adaption in the Digital Twin during the operational phase. As the knowledge about the models usually is held by the model vendor, the plant operator is faced with the challenge of identifying a suitable model for his current scenario so that the Digital Twin can represent an appropriate level of realism for the application. Therefore, a concept is needed that enables cost-effective, objective and error-free model adaption in the Digital Twin during operational phase. (Dittler et al., 2022a)

The remainder of this contribution is structured as follows. In section 2, an overview of related work is given first. Section 3 initially describes the general overview and then the concept for the model adaption. Finally, section 4 illustrates the implementation and the benefits of the concept by using an application scenario.

## 2. BASICS AND RELATED WORK

In this section, we first discuss the term Digital Twin and its extension to include intelligence. Subsequently, approaches from the literature are listed that show starting points for the implementation of an application-oriented model adaption in the Digital Twin.

*2.1 Digital Twin*

The Digital Twin (DT) is used in a wide variety of areas, such as manufacturing, healthcare and smart cities (Fuller et al., 2020; Tao et al., 2018). The DT concept, which can be traced back to its origins in the 2000s, has seen a significant increase in publications since 2016 (Sjarov et al., 2020). Over the years, a variety of definitions have emerged for the term and concept (Negri et al., 2017). The authors of (Ashtari et al., 2019) conducted a literature review and then proposed a definition for a DT that combines the essential aspects of the reviewed definitions.

The basic building blocks of such a DT are models and their relation to each other, which represent a virtual image of an asset. Furthermore, the DT must have three additional properties (Braun et al., 2022; Dittler et al., 2022b):

- The models and their relations are synchronised with the asset so that at any time the static image of the asset is replicated.
- There must be an active data acquisition from the asset to the DT so that the dynamic processes are also perceived.
- The existence of an executable model is necessary to enrich the static image with the replication of the dynamic behaviour of the asset.

If a DT is extended by a 'DT model comprehension', 'intelligent algorithms', 'services' and a 'feedback interface', it is defined as an intelligent DT (iDT) (Ashtari et al., 2019). This enables, for example, high-level functionalities such as optimisation of the process flow or predictive maintenance based on the operating data stored in the DT at runtime. The intelligence of the iDT can have multiple manifestations in this context, one of which, the fully automatic model adaption, equips the iDT with the ability to maintain application-appropriate realism during the operational phase (Dittler et al., 2022a). Agents are particularly suitable for making such an adaption automatically, which is why related approaches will be described in the following.

*2.2 Agent-based approaches*

In (Vogel-Heuser et al., 2021), the use of a Multi Agent System (MAS) to implement an Asset Administration Shell (AAS) is illustrated. The AAS as one of the central Industry 4.0 concepts realises a standardised envelope of all relevant data for a specific asset, including the management of its sub-models, which form a substantial part of the shell. In (Sakurada et al., 2022), a similar approach is proposed to orchestrate components like models and services within the AAS as well as the external communication via an agent-based interface.

In (Jung et al., 2020) the authors present a co-simulation framework that allows the integration of heterogeneous models by coupling different simulation tools within a simulation environment at runtime. The agent-based approach allows dynamics at runtime, in the sense of 'plug-and-simulate' and also the integration of non-standardised models. In (Härle et al., 2020) the authors present an agent-based assistance system for the automatic composition and configuration of a co-simulation for the commissioning phase. This is extended by the authors of (Härle et al., 2021) through the operation-parallel model adaption, which solves a modelled multi-objective optimization problem with the help of an evolutionary algorithm.

In (Müller et al., 2022), a reconfiguration management architecture for generating alternative plant configurations with a dynamic simulation model generation based on production requirements is presented. Simulation-based multi-objective optimisation is used to determine configurations that are close to the optimum. Agents are used for autonomous generation, optimisation, selection and evaluation.

The approaches shown indicate that agents are increasingly used for automated dynamic adaption in different areas and are becoming more and more relevant, especially with the trend towards decentralisation. In the context of an application-oriented model adaption in a heterogeneous model landscape - with behavioural models in different modelling depths - in the DT, the authors could not yet find any approaches in the literature. This may also be related to the fact that the automated creation of behaviour models by component vendors is also focus of current research (Stegmaier et al., 2022b). To use these models efficiently in the future, a concept for automated selection of the appropriate modelling depth is required, which is described below.

## 3. PROPOSED APPROACH

This section first gives a general overview (Fig. 1) of the functionalities (grey) required for the model adaption and the focus of the contribution (blue). Then the underlying method of the concept is described in more detail.

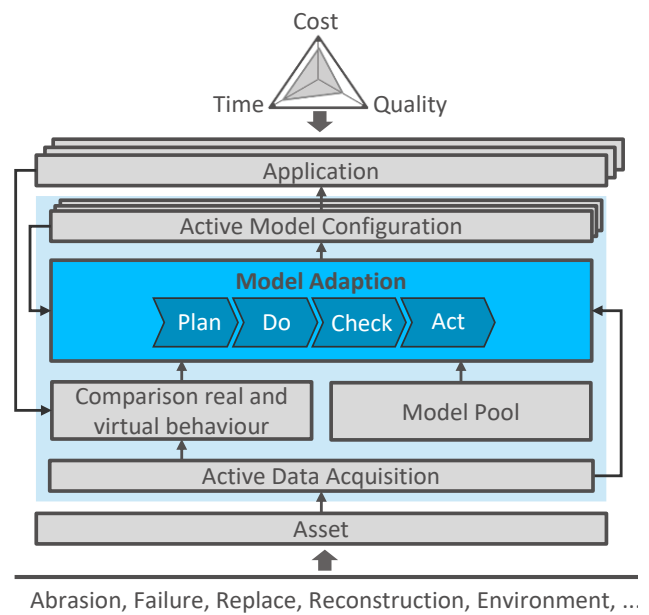

Fig. 1 Overview of the Model Adaption Approach

A new application, changes in the requirements of an application, as well as changes at the asset or its environment can lead to a so-called model drift. This opens a reality gap, which denotes a discrepancy between model and real condition. It is possible to compare the real and the virtually modelled behaviour by comparing the values supplied by a simulation with the actual values from the *Active Data Acquisition*. An identified deviation, e.g. due to a change in the environment, triggers the *Model Adaption* to provide an optimised model configuration with regard to the 'magic triangle' of time, cost and quality. This automatically provide the most suitable model configuration in the sense of a 'model-as-a-service' approach. An existing application (e.g., prognosis, optimisation or operation-parallel simulation) is supplied with the *Active Model Configuration*. To optimise the current configuration the detected deviation, the Active Model Configuration as well as the process data from the Active Data Acquisition are made available to the Model Adaption. The Model Adaption accesses the existing *Model Pool* to provide application-appropriate models.

As described in section 1 and 2, this model pool consists of a heterogeneous model landscape which needs to be structured to ease an automatic selection of the models. There are many proposals for this structuring in the literature. (Stegmaier et al., 2022a) provides a study and subsequently described a promising four-dimensional structuring of behaviour models. This structure allows models to be selected according to their application via model width, model range, model depth and behaviour type. The focus lies on the classification of model depths, which are divided into five levels, from discrete behaviour (model depth 1) to physical-spatial behaviour (model depth 5).

To counter the ageing process, the Plan-Do-Check-Act cycle, a systematic approach for a continuous improvement from Lean Management theory, has become established in industry. This is used as the basis for the development of a method for automatic model adaption. The specific manifestations of the four steps of the Plan-Do-Check-Act cycle are shown in Fig. 2 and are described in more detail in the following.

**(1) Plan:** In the first step, a possible need for adaption is determined. The three different triggers: change of asset values, change of requirements of an application or a new application are checked. The requirements of the applications are compared with the meta-information of the active model configurations with regard to functional suitability in order to determine a need for adaption. If a deviation occurs despite successful mapping of the application requirements to one of the active model configurations, it is always a change to the asset or its environment. If none of the active model configurations can resolve the identified deviation, the adaption target is set and the Do step described below is triggered. The adaption goal consists of the functional suitability and the weighting of the 'magic triangle'.

**(2) Do:** In the Do step, suitable model configurations are generated. The generation of new suitable model configurations is based on a two-staged model adaption approach. The first stage uses a minimally invasive approach, in the sense of effort reduction, to examine whether a currently active model configuration can be transferred via parameter adaption (parameterisation) into a suitable model configuration that fulfils the adaption goal. If this is not possible, the second stage is activated, in which new model configurations are orchestrated from the model pool in the form of executable simulation models. First, all functionally suitable model configurations are formed in the sense of a structural adjustment. This includes, for example, the selection of a model in a higher model depth, which can represent a detailing of the model through a finer-granular parameterisation.

**(3) Check:** In the check step, the model configurations found are compared with regard to the 'magic triangle' and then the most suitable model configuration is selected. For this purpose, the individual model configurations are simulated automatically. Based on the information about model depth, model range and model width from the model pool, a simulation-based evaluation can be carried out with regard to time, cost and quality for each corresponding application. The comparison of the values provided by the model configurations with the process data from the Active Data Acquisition enables the evaluation of the quality in terms of realism. If no adequate model configuration is found, the Do step is triggered again. If a suitable model configuration is identified, the Act step is triggered.

**(4) Act:** The Act step activates the most suitable model configuration and passes its information to the Plan step, which waits for the next trigger for a new run of the Plan-Do-Check-Act cycle. The iterative procedure with the feedback loops ensures that the model configurations are provided as accurately as necessary and not as accurately as possible, which corresponds to the basic idea of the Plan-Do-Check-Act cycle, i.e. continuous improvement.

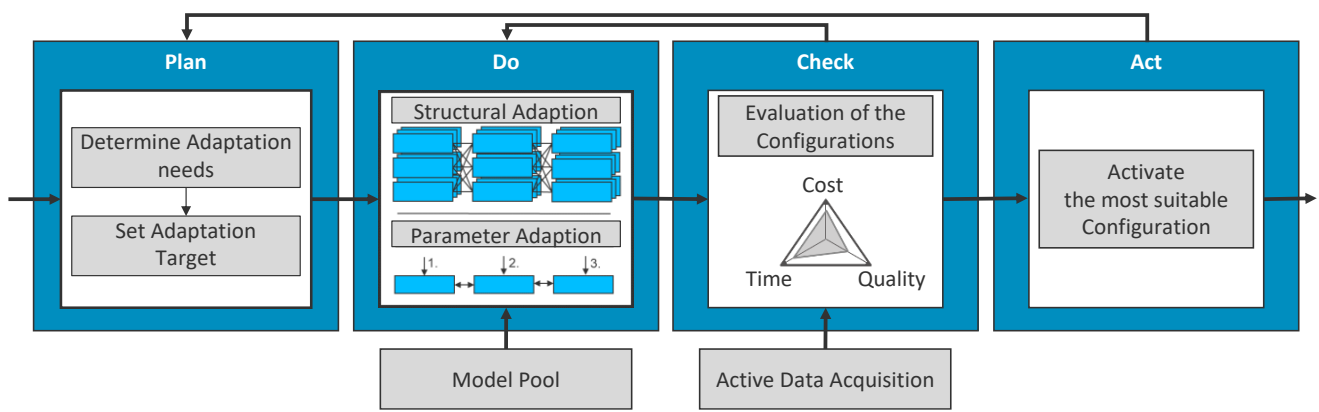

Fig. 2 Model Adaption Concept based on Plan-Do-Check-Act Cycle

In principle, this method can also be applied manually, but this can be cost-intensive, non-objective and error-prone due to the large amount of knowledge required about the heterogeneous model landscape and the number of possible configurations. Therefore, the model adaption should run autonomously. This is achieved by each step interacting with the previous and next steps and doing its task proactively and in a target-oriented manner. Due to the number of different components for the continuous model adaption, an increased communication effort arises, which can be significantly reduced by a proactive persistence attitude of the components. An agent-based system architecture well suited to fulfil these properties. Therefore, a convenient realisation is presented in section 4.

## 4. ILLUSTRATIVE EXAMPLE AND REALISATION

In this section, an application scenario is first described and then the realisation is presented. Subsequently the results of the prototype are shown, as well as the benefits provided by an agent-based realisation of the model adaption concept. Finally, an architecture is proposed that illustrates the integration of the model adaption system as a service for the DT based on a prototypical modular production system.

*4.1 Application Scenario*

The benefits of the concept are illustrated by an application scenario using the modular production system (Fig. 3, left part) of the Institute for Industrial Automation and Software Engineering. It consists of a warehouse, five processing modules and four automated guided vehicles (AGVs). The AGVs handle the transport between the processing modules, depending on the work plan of the products. The processing modules are equipped with conveyor belts to transport the products delivered by the AGVs to the processing stations of the processing modules. For the application scenario, a processing module is extended by a physical gripping system (Fig. 3, right part) as processing station. The application that uses the DT is an operation-parallel simulation to map the real system behaviour and provide the plant operator with a visualisation. This is used e.g., to monitor the productivity of the system. For this application, the models of the system components (AGV, processing modules and their processing stations) must be coupled to create an active model configuration mapping the material flow. For each system component are different models available in the model pool. The active model configuration of the application scenario is initially specified to select model depth 2 for the system components, which according to (Stegmaier et al., 2022a) corresponds to discrete-time modelling and was realised with dead-time elements in MATLAB (Dittler et al., 2022a).

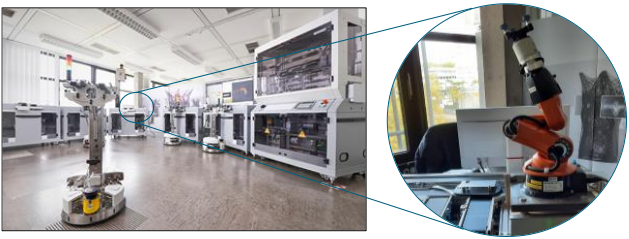

Fig. 3 Modular Production System and Gripping System

Even with this limited number of assets involved in the operation-parallel simulation of the modular production system, the increasing size of a model pool that needs to be processed within the DT becomes clear. Each component comprises a multitude of possible changes which can lead to a reality gap within the simulation. These changes are included as parameters progressively added with increasing model depth as demonstrated in (Stegmaier et al., 2022a). To determine the correct origin of the reality gap both the appropriate model and parameters for a present application must be identified.

As preparation of the scenario, the operation-parallel simulation of the vacuum gripper mounted on the physical gripping system was implemented. For this, different measurements containing deliberate alterations were prepared. Each alteration represents a possible source of error that affects the functionality of the gripper. Within this scope the model adaption system can then be used to support the diagnosis process. For further explanation, a drop-in suction pressure generated by the vacuum ejector is used as a change that often occurs in practice. This leads to a decreased contact pressure between the gripping system and the gripped work piece. In the worst-case work pieces cannot be safely held due to this insufficient pressure, which leads to damaged or destroyed components, if they are dropped. This error source should be found as quickly as possible.

The conventional manual troubleshooting by maintenance staff could lead to elongated downtimes of the machine, and thus high costs. This currently manual performed process is described in the following and will be compared to the automated model adaption system in section 4.2. In reality experienced service personnel is mostly not available directly on site and the equipment vendor has to be contacted, which leads to even higher downtimes.

Using a DT, the plant operator is provided with several simulations to enable parallel simulation-based troubleshooting. For a manual diagnosis two possibilities within this process were considered: In the best-case scenario a diagnosis expert with sufficient knowledge about the used simulations is directly available to start the troubleshooting. Relevant parameters to localise the source of error are considered right at the start of the diagnosis and the result can be verified after a short amount of time. In the worst-case scenario no expert knowledge about the simulation models is available. At first, the exact model to be used for the diagnosis must be identified before the relevant parameters can be searched and be validated. This also results in a long maintenance time. The next section describes how this manual process can be automated by the model adaption system.

*4.2 Implementation of the model adaption system*

In contrast to manual troubleshooting, the model adaption system provides a user interface that allows even less trained personnel to easily control the diagnosis. For each integrated model category, the user can pick an adaption goal (e.g., increase or decrease model depth) and a weighting factor to balance the goals of the 'magical triangle'. In the present application, the goal was to increase the model depth of the simulation, with a strong focus on simulation quality to close the reality gap. The system automatically adapts the current simulation, which can be used by a plant operator afterwards to identify the source of the reality gap and understand the problems in the physical system.

In the current state of the implementation, a JADE-based MAS realizes the concept illustrated in section 3. The single steps of the described Plan-Do-Check-Act process are realised with JADE-Agents, with each logical step realised through a single agent.

The computational logic of the model adaption system is supplemented by several partial-model agents, each responsible for a single model of the asset and containing the corresponding metadata to support the adaption. While future iterations of the system envision modules closely linked to the agent concept, e.g., the implementation of a knowledge database to increase adaption quality, in the current version MAS solely serve as an implementation paradigm. However, this already leads to a wide range of advantages.

Using the agent standard increases the interoperability with overarching concepts as well as the ease of integration in agent-based, decentralized DT applications. Besides that, a MAS provides a high extendibility, integration and swapping of components, which is even more beneficial in the current development state of the model adaption system. With a MAS agent types with higher logic like trainable algorithms can be added easily later. Further, the generic functionality of the adaption system, can be used to manage a highly diverse model pool. The flexibility of MAS enables the addition of new models (and model categories) during runtime to the model pool. The container solution of JADE further increases this factor, by enabling the decentralized execution of simulation models, for example on a server provided by the component vendor. It is beneficial for example, to outsource a model and the corresponding agent to a special simulation computer if the system involves complex simulations with a high computational effort.

Field test of the described scenario based on prepared measurements shows a significant decrease of diagnosis time when using the implemented model adaption system, as shown in Fig. 4. The proposed system reduces the troubleshooting time from 38 min (manual adaption - best case) and 96 min (manual adaption - worst case) to 33 min (automatic adaption - worst case) and 8 min (automatic adaption - best case) with the model adaption system.

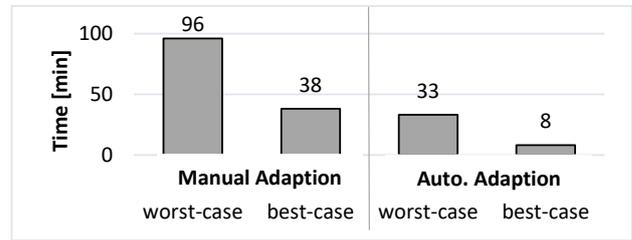

Fig. 4 Manual and automated Model Adaption

The tests performed have shown that the adaption time is strongly correlated with the simulation time of the most complex model and the number of iterations required during parameter estimation to close the reality gap, which shows the automated best case when directly finding the appropriate parameter set. In future iterations the time needed is expected to drop even further, for example by integrating an automated status monitoring system to start the process immediately. The scenarios for manual troubleshooting do not include arrival and setup time before the diagnosis can be started, which is highly optimistic. To that end, it cannot be assumed that qualified personnel with expert knowledge about the models is always available. In reality this assumption is satisfied in few cases and without experts the troubleshooting time has to be increased even more. Lastly, automating the diagnosis increases the overall reliability of the process by minimizing the chance of misconception.

In order to fully realise the concept in terms of DT, a system architecture (Fig. 5) is developed. On the *Integration & Communication* and the *Information* level, OPC UA is used to implement an Active Data Acquisition. This data is provided to the *Comparison of Real Virtual Behaviour* module to identify the reality gaps, which are passed on to the MAS.

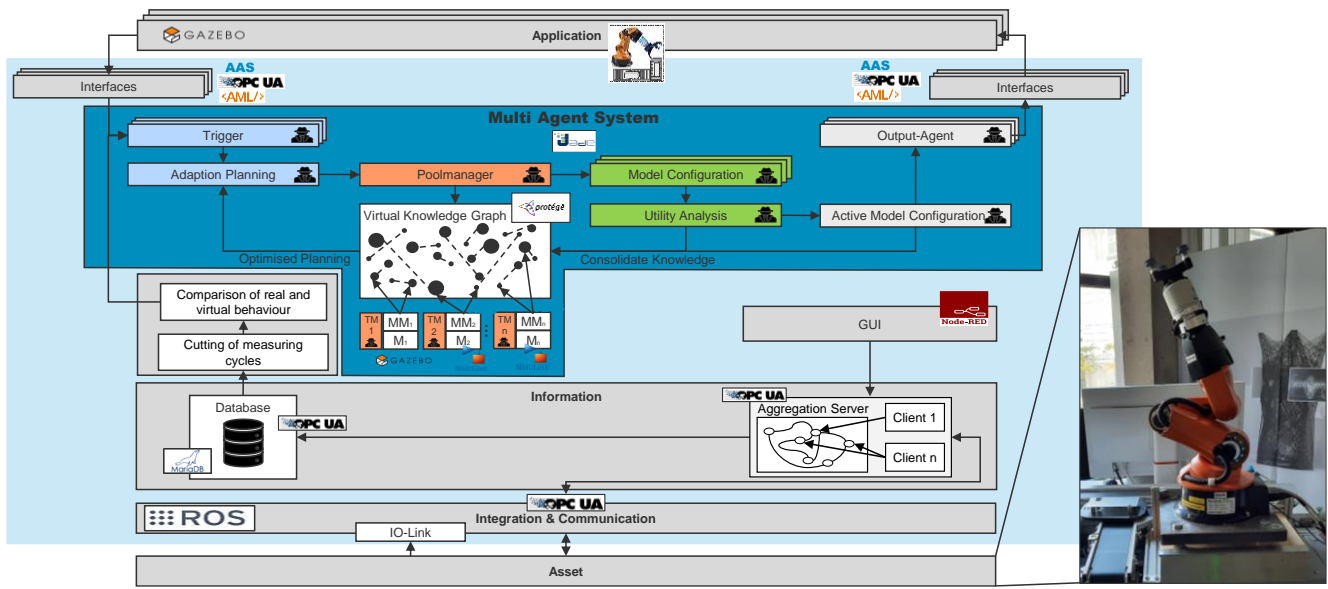

Fig. 5 System Architecture from Asset to Application

Trigger agents serve as an interface to trigger an adaption process based on the possible changes (see section 3). The Plan (blue, Fig. 5), Do (orange), Check (green) and Act (grey) method is realised according to the described concept (section 3). As described in section 4.2, the knowledge is currently mapped via partial-model agents. In the future, a Virtual Knowledge Graph will be realised that maps meta-knowledge about models such as the coupling of certain sub-models or consolidated knowledge from the Check and Act steps (e.g. scenario-specific configuration weighting) in order to further optimise the method. The corresponding model configurations are then made available to the applications (e.g. the simulation of the overall system in Gazebo) by output agents.

## 5. SUMMARY AND CONCLUSION

As this contribution showed, Digital Twins can serve as a digital 'playground' for different applications. In order to reduce the effort of orchestrating and adapting heterogeneous models for an application, a model adaption approach in the intelligent Digital Twin is presented. Besides that, the contribution shows:

- Agents are well suitable to realise an automatic model adaption as service of an intelligent Digital Twin.
- The Plan-Do-Check-Act method can be used to structure the model adaption approach in order to find a suitable configuration solution, according to the 'magic triangle', in a short time without getting lost in possible alternatives.
- In the illustrative example, the automatic model adaption system was at least three times faster than the conventional manual model adaption.
- As a positive side effect, the approach supports the plant operator in understanding the reason for the reality gap.

In the future, the knowledge about the models and their coupling, which is necessary for the model adaption, as well as the knowledge gained through the iterative procedure of the Plan-Do-Check-Act method, will be mapped by using a Virtual Knowledge Graph in order to further improve the approach.

## ACKNOWLEDGMENT

This contribution was funded by the Federal Ministry of Education and Research (in German: BMBF – Bundesministerium für Bildung und Forschung), H$_2$Mare project PtX-Wind, grant no. 03HY302R.